\documentclass{aastex}
\newcommand{\cxo}{{\sl Chandra}}

\newcommand{\hst}{{\sl Hubble}}
\newcommand{\msun}{M$_{\odot}$}
\newcommand{\ergl}{ergs~s$^{-1}$}

\newcommand{\ha}{H$\alpha$}

\newcommand{\etal}{et al.}
\newcommand{\ngc}{NGC~5953/5954}
\shorttitle{ULX in NGC~5954}
\shortauthors{Rosado et al.}

\begin{document}


\title{Detection of X-ray elongated emission from a ultraluminous X-ray source in the interacting pair of galaxies NGC~5953/5954}

\author{
Margarita Rosado\altaffilmark{1},
Kajal~K.~Ghosh\altaffilmark{2},
and Isaura Fuentes-Carrera\altaffilmark{3,4}}

\altaffiltext{1}{Instituto de Astronom\'{\i}a--Universidad Nacional Aut\'onoma de M\'exico--Apartado Postal 70-264, 04510 M\'{e}xico D. F., M\'{e}xico}
\altaffiltext{2}{Universities Space Research Association,
NASA Marshall Space Flight Center, VP62, Huntsville, AL, USA}
\altaffiltext{3}{Instituto de Astronomia, Geof\'{\i}sica, e Ciencias Atmosf\'ericas, Departamento de Astronomia, Universidade de S\~ao Paulo, Rua do Mat\~ao 1226, Cidade Universit\'aria, 05508-900 S\ao Paulo, SP, Brazil}
\altaffiltext{4}{{ GEPI, Observatoire de Paris, CNRS, Universit\'e Paris Diderot, 5 Place Jules Janssen, 92190, Meudon, France}}

\begin{abstract}
We present radio through X-ray results of a bright ($>$ 10$^{40}$ \ergl\ in the 0.5 to 8.0~keV band) ultraluminous X-ray source (ULX), CXOU J153434.9+151149, in the starburst, interacting pair of galaxies NGC~5953/5954. \cxo\ image of this ULX shows that it is elongated. 
From HST/WFPC2/F606W data we have detected a  counterpart of the ULX system with M$_{F606W}$~$\sim$~-7.1$\pm$0.7~mag. This optical counterpart may be  either an O-type supergiant star or a young star cluster.
From our Fabry-Perot interferometric observations, we have detected  \ha~ and [NII] ($\lambda$6584 \AA) diffuse emission, with velocity gradients up to 60 km~s$^{-1}$ at  the astrometric corrected Chandra position of the ULX.
Different scenarios have been invoked as to explain
the possible nature of CXOU J153434.9+151149. Based on the observed X-ray morphology of the ULX, we determine that the inclination angle to the elongated emission will be $\sim$53$^{o}$. Beaming with this geometry from a stellar-mass black hole system  will  be inadequate to explain the observed X-ray luminosity of this ULX. Finally, we
 suggest  that  mild-beaming from a binary  black hole with mass more than  50\msun\,  associated with a young star cluster, is the most favorable scenario that describes the multiwavelength properties of this ULX.
Future observations are highly essential to determine the  nature of this rare object.

\end{abstract}
%%%%%%%%%%%%%%%%%%%%%%%%%%%%%%%%%%%%%%%%%%%%%%%%%%%%%%%%%%%%%%%%%%%%%%%%%%%%%
\section{Introduction}
Theoretically, black holes evolved from single stars have masses $<$~20~M$_{\odot}$ (Fryer \& Kalogera 2000; Heger et al. 2003).
Galactic black hole systems are observed to radiate
at luminosities of 10$^{37}$~ergs~s$^{-1}$ in X-rays and occasionally up to a few $10^{39}$~ergs~s$^{-1}$
or several times the Eddington limit (McClintock \& Remillard 2006; Remillard \& McClintock 2006). Supermassive black
 holes at the centers of galaxies, on the other hand,
 have masses 10$^{6}$~--~10$^{9}$~M$_{\odot}$  and luminosites of
10$^{41}$ to 10$^{46}$~ergs~s$^{-1}$.
Thus, the off-nuclear X-ray sources in external galaxies with observed luminosities in the range
 $\sim10^{39}$ to 10$^{41}$~ergs~s$^{-1}$  in the 0.2-10 keV band are known as Ultraluminous X-ray sources (ULXs), which can describe a possible link between 
stellar collapse and the formation of Active Galactic Nuclei (AGNs).
Even though, ULXs were first discovered in the 1980's with the Einstein Observatory (Helfand 1984, Fabbiano 1989)
and are being studied with subsequent satellites (Roberts  \& Warwick  2000; Colbert \& Ptak 2002; Swartz et al. 2004; Liu \& Mirabel 2005), their physical nature is still a subject of intense debate (for recent reviews see Fabbiano 2006; King 2006; Fabbiano \& White 2006).
Contemporary explanations for the ULX phenomena fall into different broad categories: (i) accreting intermediate-mass black hole systems (IMBHs) with masses in the range 50~--~10$^{5}$~M$_{\odot}$ (Colbert \& Mushotzky 1999; Makishima  et al. 2000; Madau  \& Rees  2001; Ebisuzaki  et al. 2001; Portegies-Zwart   \& McMillian  2002; Miller  \& Hamilton  2002; Ho, Terashima \& Okajima 2003; Mushotzky 2004; van der Marel 2004; Freitag et al. 2006), (ii) geometrically/mechanically- (King et al. 2001a; Fabrika \& Mescheryakov 2001; Fabrika 2004; Poutanen et al. 2007) or relativistically beamed or super-Eddington accretion stellar-mass black hole systems (K\"{o}rding \etal\ 2002; Georganopoulus \etal\ 2002; Abramowicz et al. 1980; Arons 1992; Gammie 1998; Begelman 2002, 2006; Grim, Gilfanov \& Sunyaev 2002), (iii) supernovae and hypernovae (Terlevich 1992; Schlegel 1995; Paczynski 1998; Wang 1999; Li 2003), (iv) supersoft sources (Swartz  et al. 2002; Kong \& Di Stefano 2005), and (v) foreground/background objects, which mimic as ULXs (Arp et al. 2004; Guti\'errez \& L\'opez-Corredoira  2007 and references therein).

By analogy with the X-ray spectra of Galactic black hole binaries, a 
multicolor disk (MCD) blackbody component with 
temperature around 100 eV (plus powerlaw) has been considered as the signature of a cool accretion disk of an 
IMBH system (Miller et al. 2003; Kaaret et al. 2003; Miller et al. 2004; 2004a; Cropper et al. 2004; Kong et al. 2004). In addition,
quasi-periodic oscillations (QPOs) with long quasi periods have been detected 
in some ULXs (Liu et al. 2005; Strohmayer \& Mushotzky  2003; Soria et al. 2004; Strohmayer et al. 2007). Comparison of 
these results with the scaling relation between black hole mass and break-frequency of QPO  suggests the
presence of IMBHs in these ULXs (Belloni \& Hasinger 1990). However, the  MCD model may have different explanations in the frame works of stellar-mass black 
hole system (King \& Pounds 2003;  Roberts et al. 2005; Stobbart et al. 2006; Goncalves \& Soria 2006; Barnard et al. 2007). 
In addition, more detailed analysis of non-LTE accretion flows around IMBHs shows that the effects of black hole rotation and Compton scattering can easily generate hot accretion disks with temperatures up to kT$\sim$1~keV (Hui, Krolik \& Hubeny 2005), which is in sharp contrast to the conventional cool disk model. On the other hand, periodic variations in the X-ray light curves of some ULXs suggest that they are stellar-mass black hole binaries (Sugiho et al. 2001; Bauer et al. 2001; Liu et al. 2002, 2005; Strohmayer \& Mushotzky 2003; Pietsch, Haberl \& Vogler 2003; Pietsch et al. 2004; Stobbart, Roberts \& Warwick 2004; Weisskopf et al. 2004; Soria et al. 2004; Soria \& Motch 2004; Mukai et al. 2005; Ghosh et al. 2006; Fabbiano et al. 2006). In addition,
spectral 
curvatures have  been detected in the X-ray spectra of a few luminous ULXs (Dewangan et al.  2004 and references therein), which suggest that beaming may be a possible explanation for ULX phenomenon with stellar-mass black hole binaries. However, optical photometric and spectroscopic studies have revealed ionized nebulae around some ULXs. Some of these ionized nebulae are isotropic with huge energy content, which will be difficult to explain in the frame works of beaming models. Again, some ULXs are associated with luminous supernovae/hypernovae and supersoft X-ray sources.

All these results indicate that ULXs are not a homogeneous set of physical objects (Feng \& Kaaret 2005).
Similar studies were carried out by Winter et al. (2006, 2007) and they  classified ULXs into low/hard and high/soft states. However, long term monitoring of some ULXs displayed opposite results to what has been observed in Galactic X-ray binaries during their state transitions (Feng \& Kaaret 2006; Roberts et al. 2006).
In summary, no clear picture is emerging from X-ray studies. It is clearly evident that 
X-rays alone cannot differentiate between compact accretor models of ULXs and, in particular, cannot easily identify candidate IMBHs. Thus, the studies of ULXs in other wave bands are essential. 

Based on the results of optical imaging photometric studies, stars, star clusters and ionized nebulae  have been detected as the possible counterparts of some ULXs (Ghosh et al. 2001, 2005; Soria et al. 2005; Ptak et al. 2006; Ramsey et al. 2006). Optical spectroscopic studies on the local environments of some ULXs have also been carried out and different emission lines have been detected.  It has been suggested that both shock excitations and photo-ionization processes are responsible for the formation of these emission lines. These results suggest that both stellar-mass black holes with beaming and IMBHs with strong radiation fields may dominate shock- and photo-ionization processes, respectively, in ULXs (Abolmasov 2007a,b; Mucciarelli et al. 2005, 2007; Zepf et al. 2007).

Recent surveys of ULXs have shown that the number of ULXs correlates with the star formation rates or far-infrared luminosities of the host galaxies (Swartz et al. 2004).
During the process of interaction of two disc galaxies, gas transfer between
the galaxies can cause strong star formation and multiple supernova explosions, which may trigger the formation of ULXs.

A luminous X-ray source has been discovered in the NGC~5953/5954 system.
NGC~5953/5954 is an interacting pair of galaxies with high infrared luminosity (L$_{IR}$$\sim$$\>$10$^{44}$~ergs~s$^{-1}$, with an estimated total star formation rate (SFR)  $>$4.5 \msun ~yr$^{-1}$; Kennicutt 1998).
Given the high SFR of this system and the correlation between the SFR of a galaxy and the number of ULXs hosted by that galaxy (Swartz et al. 2004), there should be, at least 3 to 4 ULXs in this galaxy pair.
However, we have detected only one ULX, which is above the average luminosity (L$_{X} >$10$^{40}$~ergs~s$^{-1}$).
Thus, the expected ULX population of this galaxy is not present  (Gilfanov et al. 2003).
This indicates that either the missing ULXs are highly variable or they were below the detection limit at 2.7 $\times$ 10$^{39}$~ergs~s$^{-1}$ for 10 source counts.
Here we present multiwavelength results to study and explore the possible nature of the detected ULX. Multiwavelength observations, data analysis and results are presented in \S2. Discussion and conclusions are described in \S3.
We adopt a cosmology  H$_{0}$=73~km~s$^{-1}$~Mpc$^{-1}$,  $\Omega_{M}$=0.24, $\Omega_{\Lambda}$ =0.76 (Spergel et al. 2007). The cosmology corrected luminosity and angular-size distances are 28.9~Mpc and  28.5~Mpc, respectively.

\section{Multiwavelength observations and results}

NGC~5953/5954, is a binary system with an (S0/a~+~Scd) pair of galaxies (KPG~468 from Karachentsev 1972), which are transferring material (Domingue et al. 2003; Hern\'{a}ndez-Toledo et al. 2003).
Both galaxies host an active nucleus: NGC~5953 is a Seyfert~2 and NGC~5954 is a LINER.
These two galaxies are separated by a projected distance of 6.2 kpc (Hern\'{a}ndez-Toledo et al. 2003).

\subsection{X-ray observations}

NGC~5953/5954 was observed
 with the \cxo\
 Advanced CCD Imaging Spectrometer (ACIS) operating in imaging mode on both
  June 11, 2002 (ObsID 2930) and December 29,  2002 (ObsID 4023) for 9.9 and 4.7 ks, respectively.
These two datasets were retrieved from the \cxo\ archive
 and the Level~2 event lists were used to extract the events within the
 $D_{25}$ isophotes of each galaxy.
Both galaxies were fully within the back-illuminated CCD S3 field of view.
Source detection and source and background light curves and spectra were
 extracted using the locally-developed software package {\tt lextrct}
 (see Tennant  2006).
   The entire data set was then cleaned of bad pixels and
columns and the standard grade set and events in pulse
invariant (PI) channels corresponding to 0.3 to 8.0 keV
were selected for source detection. 
   No periods of high particle background occurred during
the observation.

   A source-finding method is based on a technique, which use a circular-Gaussian approximation to a known point spread function (PSF). This method  assumes that a source
is located at a given position and compares the distribution
of detected events to the PSF.
The algorithm first calculates the fraction of the PSF within
each pixel within a detection region. Then, using the PSF
fraction as the independent variable, it calculates an
unweighted least-squares fit of a straight line to the counts
detected in the pixels in the region. If a source is present,
then the slope of the line will be positive and will represent
the total number of counts from the source (integrated over
the PSF). The line intercept is the background per pixel. A
key value is the uncertainty in the slope and hence the number of source counts. The uncertainty is determined by
applying the standard propagation of errors directly to the
sums. The algorithm then calculates the estimated source counts
(slope) and error at every pixel in the image. The estimated
source counts divided by the uncertainty is the signal-to-noise ratio (S/N). If the S/N exceeds some threshold, then
there is a source in the neighborhood. This threshold is as low as 2.5
for a source on axis and as high as 3.0 for a source far off-axis. 
For the present analysis a constant value of 3.0 is used.

%-----------------------------Figure Start--------------------------------

\begin{figure}
\includegraphics[angle=-90,width=\columnwidth]{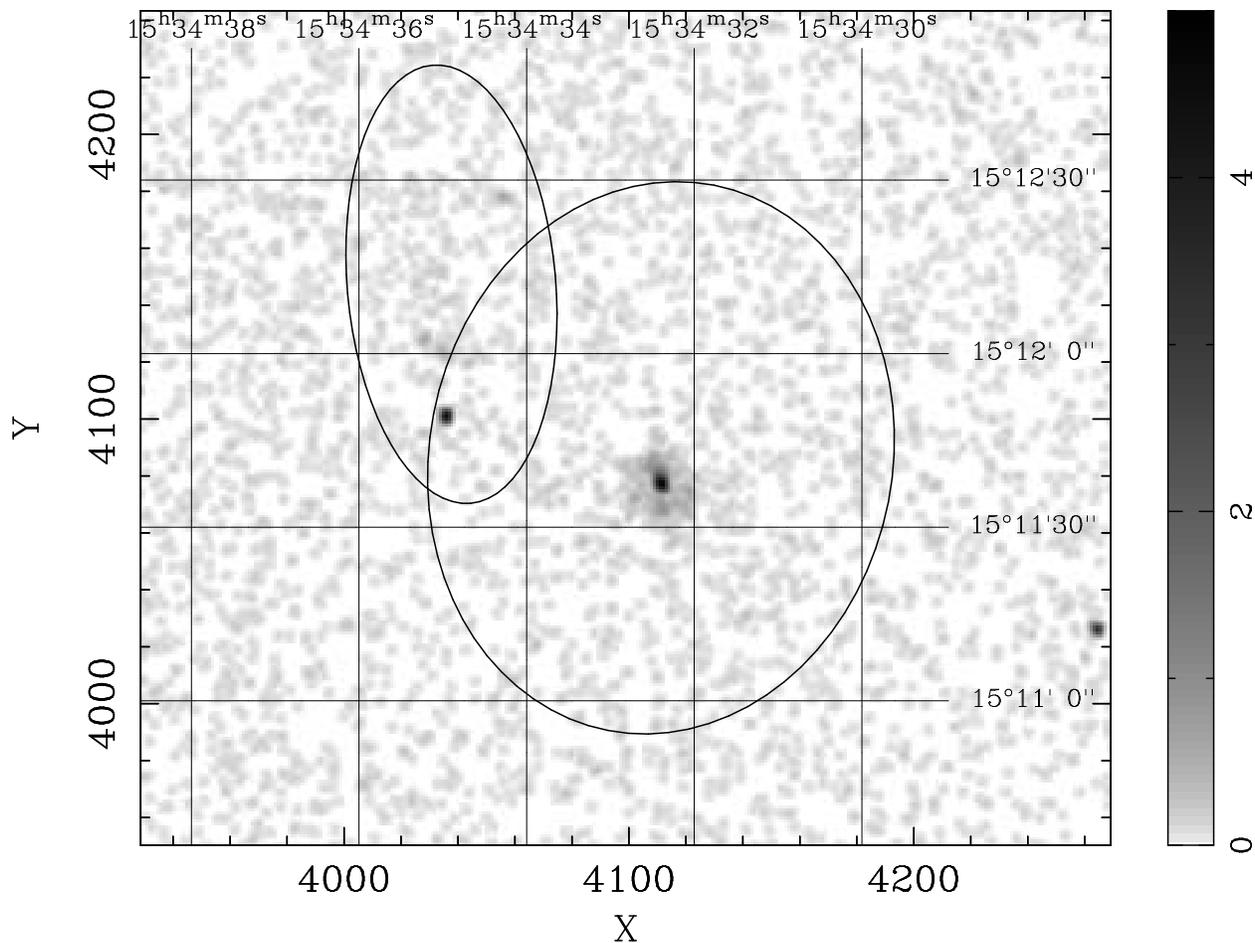}
\figcaption{The \cxo/ACIS-S3 image of  NGC~5953/5954 obtained on 2002 June 11 with $D_{25}$ isophotes drawn on the respective galaxies.
The galaxy with a bright nucleus surrounded with diffuse emission is NGC~5953.
The ULX is located in between the two galaxies.}
\end{figure}

%-----------------------------Figure End----------------------------------

We have detected two sources within the $D_{25}$ isophotes of both the galaxies in each observation.
One is the nucleus of NGC~5953 and the other is the ULX.
The detection limits for point sources in these two observations
 are $\sim$$5.9\times 10^{38}$ and $\sim$$1.2\times 10^{40}$ \ergl, respectively, in the 0.5 to 8.0~keV band.
An absorbed power-law of photon index $\Gamma=1.8$,
 the Galactic absorption column along the line of sight
 ($N_H$$=$$3.26\times 10^{20}$~cm$^{-2}$),
 and a 10~counts detection limit were assumed to compute these luminosities.
Fig. 1 shows the \cxo\ image of NGC~5953/5954 system
 with $D_{25}$ ellipses shown around each galaxy of the pair.
A elliptical Gaussian profile was fitted to the spatial distribution of X-ray events, to determine the position of the ULX, which is 
15$^{h}$34$^{m}$34.96$^{s}$~+15$\arcdeg$11$\arcmin$49.3$\arcsec$.
Following the CXO convention this ULX will be designated as CXOU J153434.9+151149.
The estimated statistical uncertainty in the ULX position is 0.\arcsec05.

Fig. 2 shows close-ups of the X-ray images of the ULX and of a point source located far from the center of the field of the observations.
As one can see, the point source 1.$\arcmin$5 from the optical axis has a symmetric image (center panel of Fig. 2).
On the other hand, the ULX image (nearer to the center of the field, where the image quality is better) appears elongated in 
the N-S direction. An ellipse with major- and minor-axis diameters $\sim$2.\arcsec5 and $\sim$1.\arcsec5, respectively, is also shown enclosing the elongated nature of the source in the left panel of Fig. 2. In addition, we have compared
the radial profile of the ULX with the PSF (model computed using the CIAO
tool {\tt mkpsf}) and is shown in the right panel of Fig. 2. It can be seen from this figure that these two profiles agree well up to 1.$\arcsec$0. However, the radial profile of the ULX clearly shows that it is broader beyond  1.$\arcsec$0 with respect to the PSF. This also displays that the ULX is extended at least up to 4 pixels ($\sim$2.$\arcsec$0).
All these results suggest that the ULX is not a point-like, but an extended source.
Indeed we can suggest that the ULX image may consist of two point sources separated by 1.$\arcsec$0 (140 pc at the NGC~5953/5954 distance). 
However, it may be mentioned that the radial profile of this ULX clearly displays gradual decline of the normalized counts per pixel.
Disabling pixel randomization may slightly improve the resolution of on-axis Chandra sources. However, in this particular case high
signal-to-noise ratio data is more crucial than the resolution. In addition, we will show in the next section that the probability of two
sources to be present within a separation of 1\arcsec.0, by random chance, is less than 3.5$\times$10$^{-6}$. 
Thus, the source is most likely an elongated object,  eventhough two sources hypothesis can not be completely ruled 
out with the existing data.

%-----------------------------Figure Start--------------------------------

\begin{center}
\includegraphics[width=4.5cm,angle=-90]{f2a.eps}
\includegraphics[width=4.5cm,angle=-90]{f2b.eps}
\includegraphics[width=4.5cm,angle=-90]{f2c.eps}
\figcaption{(left) Close-up of the X-ray image of the ULX.
The units are ACIS pixels 0.$\arcsec$492.
Note its apparent elongation along the N-S direction.
An arbitrary ellipse with major- and minor-axis diameters $\sim$2.$\arcsec$5 and $\sim$1.$\arcsec$5, respectively, has been drawn around the source and this identical ellipse will be plotted in the following figures to 
guide the eye.
(center) Close-up of the X-ray image of a point source in the field of the ULX.
Although this source is further off-axis than the ULX
(1.5$\arcmin$ vs. 0.$\arcmin$5)
its image is more point-like and symmetric. (right) The radial profile of the ULX (square) is compared with that of the PSF, which clearly shows that the radial profile of the ULX is much broader than that of the PSF.}
\end{center}
%-----------------------------Figure End----------------------------------

X-ray light curves and spectra of the ULX and the background  were extracted from a circle of radius 5\arcsec\ and a circular 
annulus of 5\arcsec\ and 10\arcsec\ radii, respectively, at the position of the ULX.
Background subtracted countrates of the ULX from the two Chandra observations were (7.5$\pm$0.8) and 
(6.6$\pm$1.3)$\times$10$^{-3}$~cts~s$^{-1}$, respectively.
This suggests that this ULX remained almost steady between the two observations.
The X-ray light curves of the ULX  were binned for
500~s and tested against the constant countrate hypothesis.
In addition, we performed the KS-test, which suggests this source did not vary during the observations.
The spectrum of the ULX for ObsID 2930 was binned to obtain at least
10~counts per fitting bin.
Spectral redistribution matrices and ancillary response
files corresponding to the ULX were generated using CIAO, v3.3.0.
XSPEC v11.3.2 was used to fit the spectrum of the ULX in 0.5 to 8.0~keV
energy band using either an absorbed powerlaw (PL) model or a
bremsstrahlung (Brem) model. Table~1 lists  the best-fitting parameters
for both spectral models.
It can be seen from this table that both models fit equally well to
this spectrum.
The Galactic hydrogen column density along that direction ($N_{\rm H}$) is only $3\times10^{20}$~cm$^{-2}$, so both
measured columns exceed this value. Bremsstrahlung temperature of 13.5 keV appears to be high and suggests for a hard source.
Similarly, the powerlaw photonindex ($\Gamma$) of 1.5 is flatter than 1.78, which is close to the average value of the powerlaw index, for all candidate ULXs with statistically-acceptable powerlaw fits (Swartz et al. 2004). This flat powerlaw index suggests for a hard source.    In addition to fitting models to the 0.5 to 8.0 keV spectrum of
the ULX, the background-subtracted X-ray
counts were binned into three broad
bands, defined as S (0.5-1.0 keV ), M (1.0-2.0 keV ), and H
(2.0-8.0 keV); and the X-ray colors (M - S )/T and (H - M )/T ,
where T = S $+$ M $+$ H, were constructed following Prestwich
et al. (2003).
X-ray color analysis also shows that this ULX is an absorbed hard source.

\begin{center}
\small{
\begin{tabular}{lccccc}
\multicolumn{6}{c}{{\sc Table 1}} \\
\multicolumn{6}{c}{ X-ray spectral parameters of the ULX} \\
\hline \hline
 \multicolumn{1}{c}{Model} & \multicolumn{1}{c}{$N_{\rm H}$$^b$} & \multicolumn{1}{c}{$\Gamma$} &
\multicolumn{1}{c}{$T_{\rm  e}$(keV)} & \multicolumn{1}{c}{$L_{\rm X}^c$} & $\chi^2$(dof) \\
% \multicolumn{1}{c}{Model} & \multicolumn{1}{c}{($10^{22}$ cm$^{-2}$)} & &\multicolumn{1}{c}%{(keV)} & \multicolumn{1}{c}{($10^{40}$~ergs~s$^{-1}$)} \\
\hline
PL$^a$         & $0.7^{+0.4}_{-0.3}$  & $1.5^{+0.8}_{-0.6}$  & --- & $1.1\pm0.2$  & 1.9(5)\\
Brem$^a$    & $0.6^{+0.4}_{-0.4}$  & ---  &$13.5^{+\infty}_{-10}$  & $1.0\pm0.2$  &  1.8(5) \\
\hline
\multicolumn{6}{l}{$^a$PL:Power-law \&  Brem: Bremsstrahlung model, $^b$in units of $10^{22}$ cm$^{-2}$}\\ \multicolumn{6}{l}{$^c$In units of $10^{40}$~ergs~s$^{-1}$in the 0.5$-$8.0 keV band.}\\
%\multicolumn{6}{l}{$^b$Brem: Bremsstrahlung model, RS: Raymond--Smith model.}\\
%\multicolumn{6}{l}{$^b$Elemental abundance is a free parameter which converges to %$Z$$=$0.}\\
%\multicolumn{6}{l}{$^c$ Cut-off energy in keV.}\\
\end{tabular}
} %end \small
\end{center}

A ROSAT HRI observation of NGC~5953/5954 was also carried out between  August 16th and 25th, 1996.
The ULX was detected with a countrate of (8.9$\pm$3.9)$\times$10$^{-3}$~cts~s$^{-1}$.
This countrate was converted into flux using the PIMMS v3.9e and the bremsstrahlung model parameters of Table~1.
The corresponding intrinsic luminosity in the 0.5-8.0 keV band is
(2.3$\pm$0.9)$\times$10$^{40}$~\ergl.
This result indicates that either the X-ray emission from the ULX has marginally decreased between the ROSAT and the Chandra observations or remained steady within the observational uncertainties.

\subsection {Hubble WFPC2 observations}

\hst/WFPC2 observations of \ngc\ were carried  
with   F218W and F606W filters on September 23, 1996 and June 10, 1994 for 1200~s and 500~s, respectively.
We could not use the data from F218W filter, because we could not convincingly determine astrometric objects using  USNO stars, 
\cxo\ and \hst\ images.
Astrometry between the \cxo\ and HST/F606W (central wavelength and halfwidth are 5997 and 1502 \AA\, respectively and thus enclosing both the redshifted H$\alpha$ and [NII] line-emission of the galaxy pair) images was performed using the nucleus of NGC~5953. It was used to register the two images 
assuming that the F606W and X-ray band centroids coincide.  F606W band includes redshifted H$\alpha$ and [NII]  lines (6584\AA). The uncertainties of the best-fit elliptical Gaussian to the X-ray nucleus is $\sim$0.\arcsec 05 
 and we take this to be the 
 registration uncertainty for the two \cxo\ sources (the nucleus and the ULX).
This uncertainty is combined in quadrature with the (statistical) uncertainties
 in the X-ray positions of these two sources ($\sim$0.\arcsec 09)
 to give their final positional uncertainties. These two sources were  on PC1 and WF3 CCDs, respectively, on the HST/WFPC2 image. We used the
 METRIC program\footnote{http://stsdas.stsci.edu/cgi-bin/gethelp.cgi?metric.hlp}
to determine the  relative 
positional errors ($\sim$0.\arcsec 27) between the PC1 and WF3 images (Ghosh et al. 2005). The resultant astrometric accuracy  is less than 0.\arcsec 30.
A potential optical counterpart was discovered within the resulting error
 circle at the \cxo\ position of the ULX candidate, which is shown in Fig. 3. Within this error circle  there may be multiple objects of single-pixel size. We have not considered these single-pixel size objects, because they are located at the boundary of the error circle and they may be just random fluctuations of the strong diffuse background emission.
However, we have 
selected the brightest object of 2$\times$2 pixels size, which is closest to the center of the error circle (shown with a black arrow in Fig. 3). Photometry of this object was done with {\tt lextrct} 
(Tennant 2006)
using the current values 
of PHOTFLAM and Zeropoint in the VEGAMAG system given in the 
\hst\ data handbook for WFPC2\footnote{http://www.stsci.edu/instruments/wfpc2/Wfpc2$\_$dhb/wfpc2$\_$ch52.html~\#~1902177}.
Apparent and absolute magnitudes of this object in the F606W band are 25.2$\pm$0.7 and -7.1$\pm$0.7~mag, respectively, assuming that it belongs to the galaxy pair.

%-----------------------------Figure Start--------------------------------

\begin{center}
\includegraphics[angle=-90,width=\columnwidth]{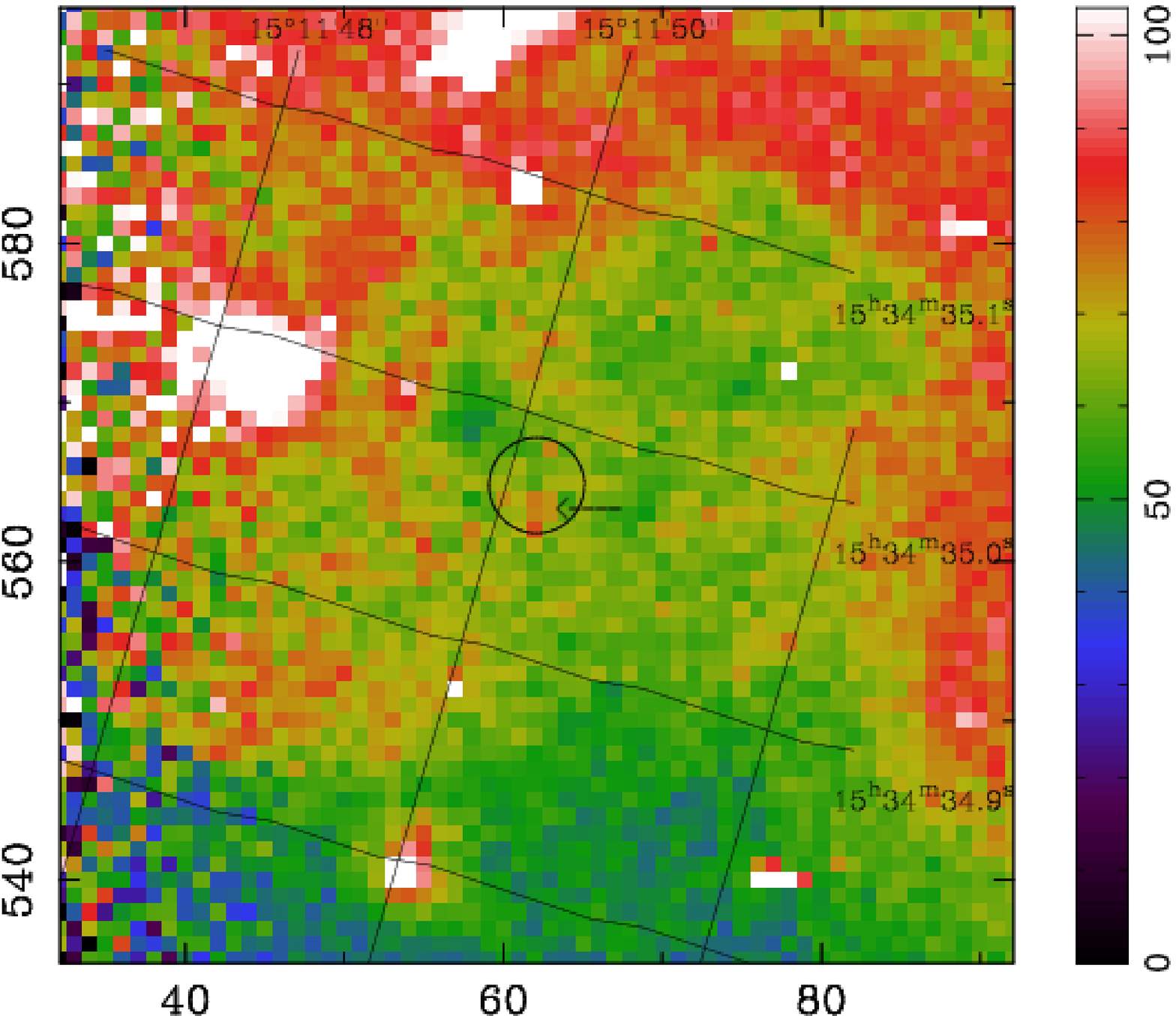}
\figcaption{HST/WFPC2/F606W image around the ULX. An error circle of radius 0.\arcsec3 is shown at the astrometric corrected \cxo\ position of the ULX. The object with four-pixel size within the circle (marked with a black arrow) is the most likely counterpart of the ULX. }
\end{center}

%-----------------------------Figure End----------------------------------

\subsection {Scanning Fabry-Perot optical interferometry}

Fabry-Perot observations of NGC~5953/5954 were carried out at the f/7.5 Cassegrain focus of the 2.1m telescope of the 
Observatorio Astron\'omico Nacional in San Pedro M\'artir, B.C., M\'exico (OAN-SPM) using the scanning Fabry-Perot interferometer 
PUMA (Rosado et al. 1995). A 1024 $\times$ 1024 Tektronics CCD detector was used with a pixel size of 0.\arcsec58.
Two Fabry-Perot (FP) velocity cubes at H$\alpha$ ($\lambda$6563 \AA) and [NII]
($\lambda$6584 \AA) were obtained on June 25th, 1998 and May 6th, 1997, respectively.
Exposure time was  2900 s for each observation.
The sampling spectral resolution is 0.41 \AA ~(19.0 km~s$^{-1}$ at H$\alpha$)   and the field of view of the H$\alpha$ and [NII] 
FP cubes is of 5$\arcmin$ and 3$\arcmin$, respectively.
Results on the [NII] kinematics of this galaxy pair derived from the [NII] data cubes and further details on the observations have 
been presented in Hern\'andez-Toledo et al. (2003).

FP data reduction and analysis were done using the ADHOCw software (developed by J. Boulesteix, http://www.oamp/adhoc/adhocw.htm) 
and the CIGALE software (Le Coarer et al. 1993).
Further description of PUMA FP observations and data reductions are described in Fuentes-Carrera et al. (2004).
The Chandra and FP images were registered using the nucleus of NGC~5953. In the case of the FP images, we also used several stars detected in the PUMA field of view in order to do the astrometry.

Fig. 4  shows a close-up of the velocity-map at +1825 km~s$^{-1}$ obtained from the H$\alpha$ FP velocity cubes.
As discussed below, the nebulosity at the ULX position has a peak near this velocity.
The ULX position is marked with an ellipse that has the same size and orientation as shown in Fig. 2.
Also marked in Fig. 4 are the  positions of two H~II nebular complexes (\#5 and \#9), which were detected by Gonz\'alez-Delgado \& P\'erez (1996).
These authors give positions as an offset from the nucleus but due
to irregularities in the emission near the center of the galaxy, the
nuclear position is uncertain at the arcsec level.
We have adjusted this fiducial point so that the two brightest HII regions
(\#16 and \#18), identified by those authors, are centered on the brighest spots in our data. This implies a shift of (2.73 $\pm$ 0.67) \arcsec between the nuclear
position given by Gonz\'alez-Delgado  \& P\'erez (1996) and the position of the
nucleus as detected in our FP images.

The ULX is located between complexes \#9 and \#5.
The sizes of complexes \#5 and \#9 are 151 pc and 165 pc, in radius, respectively, and their H$\alpha$ luminosities are 
$\sim$1.6 $\times$ 10$^{39}$ ergs~s$^{-1}$ with an estimated  rms electron density of 3.1 cm$^{-3}$ (Gonz\'alez-Delgado \& P\'erez 1996).
In our FP velocity maps, the emission of nebular complex \#9 extends
beyond its reported diameter, while for complex \#5, the emission lies
within the reported diameter.
An inspection of our H$\alpha$ and [NII] velocity-maps shows that no conspicuous point-like emission is visible at the ULX position; 
however, extended, diffuse emission is detected.
H$\alpha$ luminosity of the diffuse emission at the ULX position has been computed by extracting total counts from a circle of 
0$\arcsec$.5 radius at the Chandra position of the ULX candidate and by subtracting the equivalent background taken from annuli of 
0.$\arcsec$5 and 1.$\arcsec$0 radii of the same image, where no emission is observed.
These counts were converted into H$\alpha$ flux using the H~II complex~\#5 as the H$\alpha$ flux calibrator 
(Gonz\'alez-Delgado \& P\'erez  1996).
Computed  H$\alpha$ luminosity of the ULX is  $\sim$1 $\times$ 10$^{38}$~\ergl.
Although similar counts were obtained for the [NII] emission, implying a similar luminosity, we are unable to determine its absolute  
value, because there is no absolute flux calibrator for the [NII] line, within the field of view of our observations.
Thus, approximately, we may consider that the  [NII]($\lambda$~6584 \AA)/H$\alpha$ line-ratio will be around one, which is consistent 
with the [NII]/H$\alpha$ line-ratios of most of the HII complexes in the galaxies NGC~5954 and NGC~5953 
(Gonz\'alez-Delgado \& P\'erez  1996).

Given the diffuse appearance of the line emission at ULX position, we have extracted the velocity profiles integrated over a square 
window of 5 $\times$ 5 pixels (linear extension of 315 pc at the assumed distance) centered at the location of the ULX.
The profiles show two components whose heliocentric velocity values are confirmed from both H$\alpha$ and [NII] observations: (1) the 
brighter one at +1821 km~s$^{-1}$ (for H$\alpha$) and +1814 km~s$^{-1}$ (for [NII]) and (2) a weaker component at +1760 km~s$^{-1}$ 
(for H$\alpha$) and +1763 km~s$^{-1}$ (for [NII] at about 1$\arcsec$.5 away from the ULX location.).
Thus, within the uncertainties of $\pm$ 5~km~s$^{-1}$ in the determination of the peak velocities, both the H$\alpha$ and the [NII] 
observations show the same velocity components separated by 60$\pm$7~km~s$^{-1}$, suggesting possible expansion at a velocity of 30$\pm$7~km~s$^{-1}$ or simple 
superpositions along the line of sight of nebulae at different velocities, while the neighboring HII complexes show only a single 
velocity component at about +1800 km~s$^{-1}$.
The velocities, diffuse appearance and the [NII]/H$\alpha$ line-ratio are consistent with the idea that this ULX could be associated 
with an emission-line nebula. Emission nebulae (a few hundred parsecs in diameter) with both low- and high-ionization emission lines 
have been detected around some ULXs (Wang 2002, Pakull \& Mirioni 2003). 
Higher spatial resolution and high signal-to-noise ratio observations will be able to detect the morphology and kinematics of the 
nebulosity  at the position of the ULX.

An interesting remark regarding the galactic location of ULX is that both
ULX and the HII complexes \#5 and \#9  are located in a spiral arm of NGC~5954.
This spiral arm has a strange appearance because part of it is not curved but straight, this distortion  probably being an effect of  
tidal interaction with the companion Sa galaxy NGC~5953.
The ULX and the mentioned complexes are located where the spiral arm bends becoming straight.

%-----------------------------Figure Start--------------------------------

\begin{figure}
\includegraphics[angle=-90,width=\columnwidth]{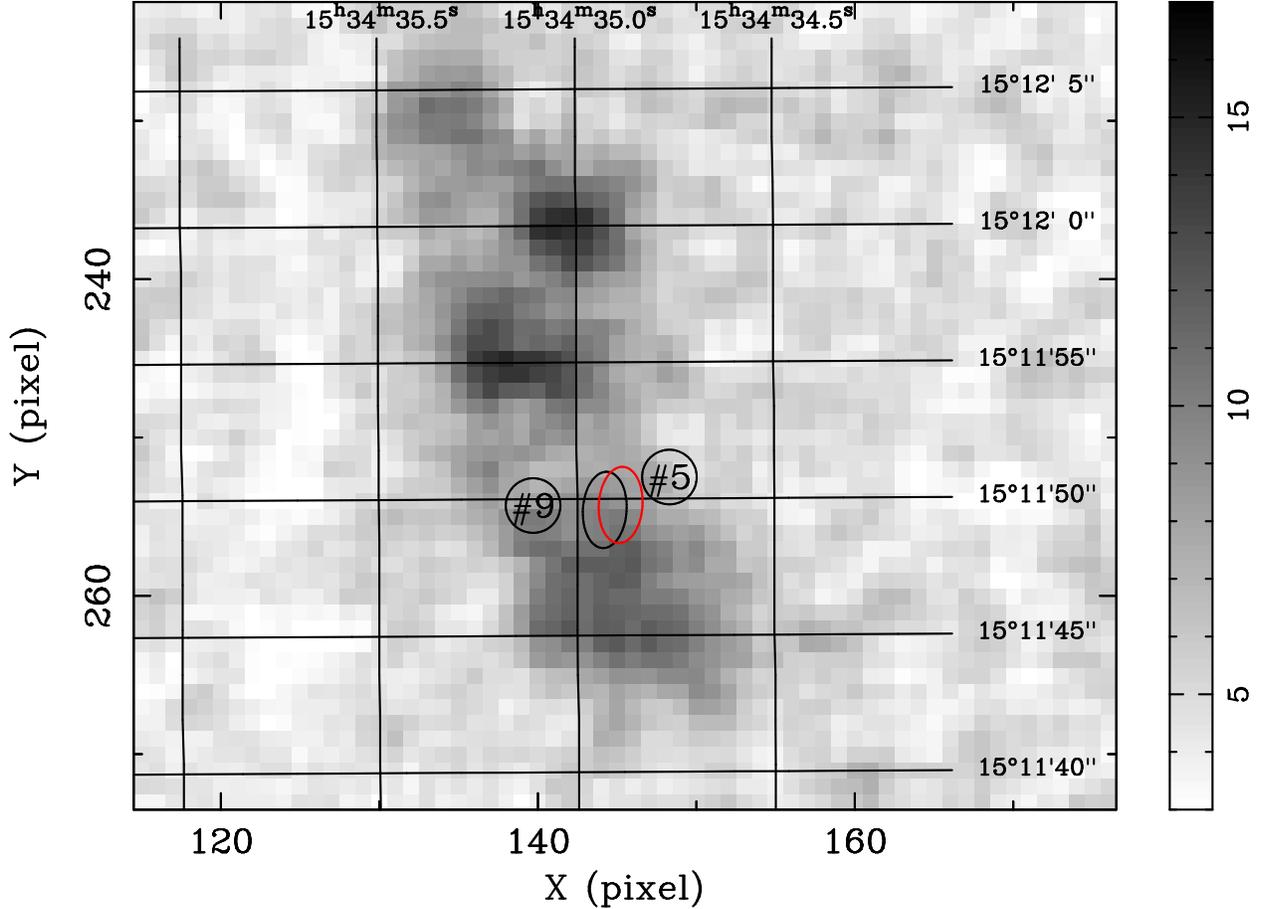}
\figcaption{Close-up of the H$\alpha$ scanning Fabry-Perot velocity-map at +1825 km s$^{-1}$ of the galaxy pair NGC~5953/5954 showing 
part of the southern spiral arm of the Scd galaxy NGC~5954.
Astrometry between the Fabry-Perot H$\alpha$ image and the Chandra was done using the nucleus of NGC~5953 (not shown in this image).
The ULX position is marked with an ellipse of same dimensions as in Fig. 2.
The black ellipse corresponds to the shifted Chandra position 
whereas the red one corresponds
to the astrometric corrected Chandra position of the ULX.
The approximate positions of the nearby HII complexes numbers 5 and 9 catalogued by Gonz\'alez-Delgado \& P\'erez (1996) are marked 
with circles whose diameter is about the dimensions reported by those authors (see the text for further details).}
\end{figure}

%-----------------------------Figure End----------------------------------

\subsection {Near-infrared and radio observations}

Near-infrared images of NGC~5953/5954 were retrieved from
the 2 Micron All Sky Survey (2MASS) archive.
Astrometry between the 2MASS  and the Chandra images was done using the nucleus of NGC~5953 (offset in R.A. is 0.00$^{s}$ and in 
Dec. 0$\arcsec$.3).
Although there are 2MASS sources in the area, they do not match peaks in our
H$\alpha$ data.
No point-like 2MASS sources were detected at the ULX position at the detection limit of 20 mag.

The NRAO Very Large Array~--~Faint Images of the Radio Sky at Twenty-centimeters (VLA--FIRST, Becker, White \& Helfand 1995) image of 
\ngc\ field
is shown in Fig.~5.
The peak flux densities at the nuclei of NGC~5953 and NGC~5954 are 16.72 and 2.95 mJy beam$^{-1}$, respectively, with an rms of 0.15 mJy.
Once again, to help guide the eye, we plot the ellipse at the Chandra position of the ULX.
This shows that no radio emission was detected at the position of the ULX during the FIRST observation although there is a weak 
radio source nearby with integrated flux density around 0.7~mJy.

%-----------------------------Figure Start--------------------------------

\begin{center}
\includegraphics[angle=-90,width=\columnwidth]{f5.eps}
\figcaption{FIRST image of  NGC~5953/5954.
The black ellipse is the same as in Fig. 2 and shows the location of the ULX candidate.
 }
\end{center}

%-----------------------------Figure End----------------------------------

\section {Discussion and Conclusions}

From \cxo\ observations of \ngc\, we have detected a bright ULX, which  is most likely an elongated object around 350 pc long.
Both powerlaw and bremsstrahlung models fit equally well to
the \cxo\ spectrum of this ULX. 
Possible optical counterpart of this ULX system is an object of absolute magnitude -(7.1$\pm$0.7)~mag. We do not know the color of this object. Since, the ULX is a bright X-ray luminous source,  we thus expect that its optical companion is
an O-type supergiant. However, the possibility of a young star cluster (Wyder, Hodge \& Cole 2000)  can not be ruled out, especially when some ULXs have been detected in the vicinity of young star clusters (  Zezas et al. 2002; Karret et al. 2004;  Krauss et al. 2005).
From 
Fabry-Perot observations, we have detected excess diffuse H$\alpha$ and [NII]~($\lambda$6584 \AA) emissions, above the ambient background, at the 
astrometric-corrected \cxo\ position of the ULX. Fabry-Perot velocity maps also show that this ULX is most likely associated 
with an emission nebula, which is expanding at a rate of 30$\pm$7~km~s$^{-1}$.  Counterparts of this ULX are not detected in 
the 2MASS and VLA/FIRST data.

Based on these results, different scenarios can be invoked to describe the possible nature of the ULX system:  (1) a foreground star, 
(2) a background object, (3) a gravitational lens system, (4) ULX consists of two ULXs, (5) an X-ray transient, (6) a supernova remnant 
and (7) an X-ray beamed system. Here we discuss these possibilities in detail.

The ratio between the X-ray and optical fluxes (F$_{X}/F_{O}$) is around 140, which is extremely large  for any stellar object. 
Value of this ratio for stars ranges between 10$^{-4}$ and 0.1 (Maccacaro et al. 1988). In addition, the value of the hydrogen column density is more than an order of magnitude higher than that of the Galactic value (Table 1). Thus, we can rule out the foreground origin 
of this ULX.

Similarly, the value of F$_{X}/F_{O}$
for normal galaxies, clusters of galaxies and AGNs is much lower than the observed value of around 140 (Maccacaro et al. 1988; 
Stocke et al. 1991). For blazars, this ratio could be up to 50. Even if we assume that the ULX is an
X-ray selected blazar with  F$_{X}/F_{O}$  $\sim$140, then the expected radio flux density at 1.4 GHz, assuming synchrotron
emission from the same population of electrons responsible for the optical emission, the expected flux density
  will be more than 50 mJy (Stocke et al. 1991; Landt et al. 2001), which is far above the emission detected from the nearby, weak radio source located at  5\arcsec\ from
the Chandra position of the ULX (Fig. 5). Thus, we can confidently rule out that the ULX is 
a background object.

If the elongated  image of the ULX is due to double sources, then it may be possible that it is a gravitational lens system. In this 
scenario, we expect two optical objects at the positions of the two X-ray sources. 
However, from the HST image we do not see a second object within the astrometric corrected error circle at the \cxo\ position of the 
so called second X-ray source. Thus, we rule out this possibility.

It may be possible that the elongated shape of the ULX is due to the presence of two ULXs separated by an arcsec. With the available 
data  we can neither  rule out nor establish this hypothesis. However, it can be mentioned that the radial profile i.e., the radial 
distribution of the counts per pixel at the position of the ULX, gradually declines. This is shown in the right panel of Fig. 2. In addition, we would like to mention that 
the chance probability of detecting two ULXs
seperated by an arcsec among 4000 sources detected in more than 100 galaxies observed with \cxo\ (Swartz et al. 2004) is extremely low, less than 3.5$\times$10$^{-6}$.
These results suggest that the \cxo\ image of CXOU J153434.9+151149 is elongated.

Extremely low values of the ratio of radio to X-ray fluxes (F$_{R}/F_{X}$) have been detected in X-ray transients 
(Fender \& Kuulkers 2001) and also in the ULX candidate in NGC~5408 (Kaaret et al. 2003). Non-detection of radio emission at the 
position of the ULX suggests that the F$_{R}/F_{X}$ for the ULX candidate in  NGC~5953/5954 is consistent with those of X-ray transients.
However, this ULX  was detected three times in three observations.  Thus, it may not be an X-ray transient. 

Extended X-ray morphology of the ULX indicate that a young supernova remnant (SNR) may explain the properties of this ULX. 
Typically, young SNRs have X-ray luminosities of $\sim$10$^{38}$ ergs~s$^{-1}$, which is two orders of magnitude lower than the 
X-ray luminosity of ULXs.
However some supernova  (SNIIn) explosions are extremely powerful, which can produce X-ray luminosities up to  10$^{41}$ ergs~s$^{-1}$ 
(Fabian \& Terlevich 1996).
These rare events occur only in late type galaxies, mainly in Sc galaxies, and are thought to be due to the explosion of a 
massive star in a very dense circumstellar medium, probably ejected in a previous evolutionary phase of the supernova progenitor,
i.e. during  its red giant phase.
These SNRs evolve  faster in comparison with the normal evolution of SNRs in less dense media
and enter into the radiative phase bypassing the Sedov phase. They  reach maximum luminosity in less than 20 years after the SN explosion, 
when their radii are small (R $<$ 0.1 pc).
H$\alpha$ luminosities and expansion velocities of these SNRs could reach up to 10$^{40}$~ergs s$^{-1}$ and   thousands of km~s$^{-1}$,
respectively. In fact, the evolution and duration of the X-ray bright phase of Type IIn SNe is very uncertain, which is detected up to 25-30 years after explosion (Fabian \& Terlevich 1996; Lenz \& Schlegel 2007).

From our FP observations we have found that the nebula around the ULX could be expanding at a rate of 30~km~s$^{-1}$,  which is too small 
compared to those of SNIIn/SNRs (Fabian \& Terlevich 1996). In addition, the major-axis diameter of the ellipse drawn on the X-ray 
image of the ULX (left panel of Fig. 2) is $\sim$2.\arcsec5, which corresponds to $\sim$350~pc. This is extremely large compared to 
$\sim$0.2~pc (Fabian \& Terlevich 1996). Thus, the observed properties of the ULX in \ngc\ are different from those of
SNIIn/SNRs.

HST/WFPC2 results suggest that the possible counterpart of the ULX in \ngc\ may be either an O-type supergiant star or a young star cluster. In addition, steady emissions have been detected from this
ULX on all three occasions in excess of 10$^{40}$~ergs s$^{-1}$. Observed, persistent X-ray emissions may originate from stable accretion disks, which are generally formed through the thermal time scale mass transfer (King et al. 2001a). It has also been shown that both stellar-mass black hole and IMBH binaries can emit persistent X-ray emissions (Kalogera et al. 2004). Thus, the luminous X-ray emission from  CXOU J153434.9+151149 can be explained in the frame works of both the stellar-mass black hole and the IMBH systems (Begelman 2006; Copperwheat et al. 2007;   Mizuno et al. 2007; Patruno \& Zampieri 2007 ). 

It is very likely that high-mass binary stars
will form in the environment of strong OB-associations on the spiral arm of the galaxy, where the ULX is located. Fast evolution
of these binary stars will lead to high-mass X-ray binaries. Thus, beaming from a stellar-mass black hole binary with an O-type supergiant star  can emit X-rays in excess of 10$^{40}$~ergs s$^{-1}$ (King et al. 2001b). If we assume that the maximum mass of a stellar-mass black hole is 20~\msun, which is accreting at the Eddington rate, then the beaming has to enhance the emission by a factor of 10 to produce the observed luminosity of the ULX. This factor of 10 can be equated with $\delta^{3+2\alpha_{x}}$, where the Doppler factor, $\delta=1/[\gamma(1-\beta\cos{\theta})]$, $\gamma$ is
the Lorentz factor of the outflow, $\beta$ is the three velocity of
the outflow in units of the velocity of light, $\alpha_{x}$  is the spectral index ($\alpha_{x}$=0.5, Table~1),
 and $\theta$ is the inclination angle to the outflow (Lind \& Blandford 1985). This gives the value of $\delta$ equals to 1.78. This value of $\delta$ can be achieved when the values of $\theta$ are 0$^{o}$, 15$^{o}$ and 30$^{o}$ and the corresponding $\beta$ values are 0.6, 0.64 and 0.9, respectively. These results demonstrate that to enhance the intrinsic X-ray intensity  by a factor of 10, the outflow has to be viewed by the observer, at least, within 30$^{o}$. However, the projected
minor-to-major axial ratio of the ellipse (Fig. 2) is roughly around $1.\arcsec5/2.\arcsec5=0.60$.
This suggests that the inclination angle  to the elongated emission is $\sim$53$^{o}$, assuming that the thickness of the elongated emission is  sufficiently small.
With this value of $\theta$, we have computed a range of values of $\delta$ for different values of $\beta$ between 0.1 and 0.9 and find that the maximum possible enhancement of the X-ray emission will be only by a factor 3.1 with the value of $\beta$ equals to 0.6. Thus, the observed X-ray morphology of CXOU J153434.9+151149, which constrains the inclination angle, indicates that the observed X-ray luminosity can not be explained with beaming from a stellar-mass black hole binary. However, changes by one pixel in the minor- or major-axis diameters of the X-ray elongated emission will lead to the inclination angle close to 30$^{o}$. Then, beaming from stellar-mass black hole binary system will be able to produce the observed X-ray brightness of CXOU J153434.9+151149 with  $\beta$ equals to 0.9. Thus, high S/N \cxo\ image of this ULX is essential to accurately determine its morphology. In addition, 
kinetic energy of such a highly relativistic outflow (velocity =0.9~c) will inflate the local
environment and will form special spatial structure. Ionized nebulae with bubble-like morphology have been detected around some ULXs (Roberts et al. 2003;
Pakull \& Mirioni 2003; Pakull, Gris\'e \& Motch 2006). These bubbles are several hundred parsec in diameter with expansion velocities around 50--80 km~s$^{-1}$ (Rosado et al. 1981, Rosado et al. 1982, Valdez-Guti\'errez et al. 2001).  It has been suggested that the  expansion could be due to the combined action of energetic supernova explosions and stellar winds, 
or to continuous inflation by geometrically-beamed jets (Chu \& Mac Low 1990, Miller 1995,   Valdez-Guti\'errez et al. 2001, Wang 2002, 
Pakull \& Mirioni 2003, Pakull, Gris\'e \& Motch 2006). Some nebulae show barrel-type shapes or enhanced 
emission along  opposite directions that could be interpreted as excitation from a beamed source (Roberts et al. 2003;
Pakull \& Mirioni 2003). However, some nebulae have displayed spur-shape, which requires an isotropic flux of energetic photons  to explain the observed HeII flux (Kaaret et al. 2004; Pakull, Grise \& Motch 2006). 
Nebulae formed by the combined action of supernova remnants and stellar winds are not supposed to have HeII emission unless it contains very massive stars.
Thus, detection of an
non-spherical HeII nebula at the position of the ULX will support the beaming model of the ULX in \ngc. 

Based on the X-ray morphology of CXOU J153434.9+151149, we have seen above that the beaming could enhance the intrinsic X-ray intensity, at the most, by a factor of three. This clearly suggests that the mass of the accretor of this ULX system has to be more than 50~\msun, assuming that it is accreting at the Eddington rate. However, the X-ray colors and spectral results indicate that this ULX is a hard/flat powerlaw source ($\alpha_{x}$=0.5, Table~1 and section 2.1), which indicates that it is a sub-Eddington system. If we assume that L$_{X}$/L$_{Edd}$$\sim$0.1, then the mass of the accretor will be more 500~\msun. This type of IMBH system can be hosted in young star cluster, which is most likely present within the error circle at the position of the ULX (shown in Fig. 3). In addition, IMBH system will photoionize its local environment, which can be detected from the optical spectra. Future observations, which we have planned to carry out, will reaveal the true nature of this system.

In conclusion, we have detected  a bright ULX in NGC~5953/5954. Its \cxo\ image is elongated.  Most likely, its optical companion is either an O-type 
supergiant star or a young star cluster. This ULX is associated with an  
extended nebular H$\alpha$ and [NII]~($\lambda$6584 \AA) emissions with a [NII]/H$\alpha$ 
line-ratio of about unity.
This nebula is moving at the characteristic velocities of the galaxy NGC~5954.
A nearby, faint radio source is located at 5\arcsec\ from the \cxo\ position of the ULX. All these results are best explained in the
frame work of a model, which consists of mildly-beaming  black hole binary of mass more than 50~\msun. 
Follow-up   radio, optical and X-ray observations are highly wanted to clearly establish the nature of this ULX.

\begin{acknowledgements}
The authors  wish to express their sincere thanks to the referee for valuable comments and critical reading of the manuscript that helped us to improve the quality of the paper. We  thank  Allyn F. Tennant for his interesting discussions on the paper.
This research has made use of the NASA/IPAC Extragalactic Database (NED);
and  the Chandra Data Archive.
Support for this research was provided in part by NASA under
 Grant NNG04GC86G issued through the Office of Space Science and from the Space Telescope Science Institute under the grant HST/AR-10954.
MR acknowledges the financial support from grants  46054-F from CONACYT and IN100606  from DGAPA-UNAM.
IFC acknowledges financial support by FAPESP fellowship 03/01625-2.

\end{acknowledgements}

\end{document}